\documentstyle[11pt,gruzinov,twoside]{article}
\begin{document}
\title{Collisionless Halos Around Black Holes}
 \author{Andrei Gruzinov}
\affil{Institute for Advanced Study, School of Natural Sciences, Princeton, NJ 08540}

\begin{abstract}
When a black hole accretes slowly, the radiative cooling of the infalling gas is weak and the accretion disk does not form. A hot collisionless quasi-spherical halo is formed instead. The properties of such halos are discussed. The rate of accretion, the radiative efficiency, and the temperature and density near the hole are evaluated. 
\end{abstract}

\section{Introduction}
In 1970, Victor Shvartsman introduced the idea of hot halos around slowly-accreting black holes. Today, we believe that these halos do exist around the supermassive black holes in the centers of non-AGN galaxies. The radio source Sgr A$^*$ is the nearest example. 

The radiative efficiency of collisionless halos is determined by the electron heating mechanisms (\S 2). The rate of accretion is not necessarily given by the Bondi formula (\S 3). The plasma density and temperature profiles near the black hole are estimated in \S 4. We conclude in \S 5.

\section{The Radiative Efficiency}
At low densities, the radiation is weak and the accreting gas is unable to cool. More precisely, the protons are unable to cool. The proton temperatures are close to $0.1{\rm GeV}$ near the last stable orbit. For realistic densities, the plasma is collisionless. Electrons synchrotron cool efficiently and are much colder than ions. Electrons radiate away nearly all the heat they receive. The radiative efficiency of accretion ($\eta \equiv L/\dot{M} c^2$, $L$ is the luminosity, $\dot{M}$ is the accretion rate) is determined by the electron heating processes.

The electron heating processes depend on the magnetic field strength. The magnetic field should be close to equipartition with the gravitational energy. Shvartsman (1971) explains this as follows. For a nearly spherical accretion, assuming perfect flux-freezing, the radial magnetic field is $\propto r^{-2}$, where $r$ is the radius. The magnetic energy density is $\epsilon _m\propto r^{-4}$. Assuming a free fall, the velocity is $\propto r^{-1/2}$. From continuity, the density is $\propto r^{-3/2}$. The gravitational energy density is $\epsilon _g\propto r^{-5/2}\ll \epsilon _m$. But the magnetic energy cannot exceed the gravitational energy, therefore $\epsilon _m\sim \epsilon _g$. 

When magnetic fields are close to equipartition, electrons should receive approximately as much heat as ions, that is about $0.1\dot{M}c^2$. There are two different mechanisms of electron heating. First, the viscous heat is evenly distributed between electrons and ions (Gruzinov 1988, Quataert \& Gruzinov 1999). Second, reconnection should be an efficient electron heater (Blandford 1998, Quataert \& Gruzinov 1999). 

If the electron heating is not negligible, as these papers suggest, the radiative efficiencies as small as $10^{-5}$ are impossible. Then the observed low luminosity of the non-AGN supermassive black holes should be explained by the small rate of accretion. Blandford \& Begelman (1998) believe that the small accretion rate is explained by the wind loses. A different explanation is proposed in \S 3.

\section{The Accretion Rate}

The rate of spherical accretion onto an object of mass $M$ is given by the Bondi formula
\begin{equation}
\dot{M}_{\rm Bondi}\sim \rho c_sR_A^2,
\end{equation}
where $\rho$ and  $c_s$ are the density and the speed of sound of the accreting gas at a large distance from the object, and $R_A=2GM/c_s^2$ is the accretion radius. For a supersonic wind, $c_s$ is replaced by the wind velocity. The Bondi formula is widely used in astronomy. 

This formula was derived assuming that the flow is: (i) spherical, (ii) inviscid, (iii) adiabatic. Although neither of these assumptions is realistic, the Bondi formula is a good order of magnitude estimate for the rate of accretion in many astronomically relevant cases (\S 3.1). We show, however, that in the case of low-density accretion onto a black hole, the Bondi formula can overestimate the rate of accretion by orders of magnitude (\S 3.2, \S 3.3) \footnote{Most of the accretion disk papers treat $\dot{M}$ as an adjustable parameter. Here we actually calculate it, just like Bondi did, but under different assumptions.}. 

For low-density quasi-spherical accretion onto a black hole, the accreting matter is a hot collisionless plasma. This plasma is an ideal heat conductor along the magnetic field lines. Heat conduction strongly violates the adiabaticity assumption used to derive the Bondi formula. 

In the absence of heat conduction, thermal energy of the plasma is lost through the event horizon. When heat conduction becomes important, some of this heat is transfered out, away from the hole. The temperature and pressure of the accretion flow increase over what have been the case for the adiabatic solution. The increased pressure opposes the gravitational attraction of the hole and impedes the accretion.

In \S 3.1 we derive the Bondi formula and explain why it works for real (non-spherical, viscous, non-adiabatic, but relatively cool) flows. In \S 3.2 we show that if the effective adiabatic index of the flow $\gamma _{\rm eff}\equiv d{\rm ln}p/d{\rm ln}\rho $ is greater than 5/3, the accretion slows down. We then show that heat conduction leads to $\gamma _{\rm eff}>5/3$, and derive a formula for the accretion rate in the presence of heat conduction. In \S 3.3 we present a more rigorous derivation of the same formula. 

When heat conduction is important, the rate of quasi-spherical accretion onto a black hole is
\begin{equation}
\dot{M}\sim\dot{M}_{\rm Bondi}\left( {R_S\over R_A}\right) ^{\alpha },
\end{equation}
where $R_S=2GM/c^2$ is the Schwarzschild radius \footnote{Relativistic effects are much smaller than the effects we are considering in this paper. We treat our black holes as Newtonian holes in space, and neglect the special-relativistic effects.}, and $\alpha \sim 1$ is a model-dependent dimensionless number.

To illustrate the astrophysical applications of the accretion rate formula (2), consider the case of Sgr A$^*$, which is believed to be a $2.5\times 10^6M_{\odot }$ black hole located at the center of our Galaxy (Genzel et al 1994, Ghez et al 1998). The accretion luminosity of Sgr A$^*$ is believed to be $L\sim 10^{37}{\rm erg}/{\rm s}$. When stellar winds in the vicinity of the galactic center collide and shock, they produce a gas of $c_s\sim 1000{\rm km}/{\rm s}$ and $\rho \sim 10^{-20}{\rm g}/{\rm cm}^3$ (Coker \& Melia 1997). The Bondi accretion rate of this gas is $\dot{M}_{\rm Bondi}\sim 10^{21} {\rm g}/{\rm s}$. Thus $\dot{M}_{\rm Bondi}c^2\sim 10^{42} {\rm erg}/{\rm s}$ is five orders of magnitude higher than the actual luminosity. We (Gruzinov 1998, Quataert \& Gruzinov 1998) and others (Shvartsman 1971, Meszaros 1975, Blandford 1998) have argued that such low radiative efficiencies are unrealistic. It is then natural to assume that Sgr A$^*$ accretes at a much smaller rate (Blandford \& Begelman 1998). If we use the estimate (2), and, for illustrative purposes, assume $\alpha = 0.4$, the estimated accretion rate is reduced by a factor of 100. The radiative efficiency is then $0.1\% $. This radiative efficiency is still small, but it might be reasonable because electrons might be heated less than ions, and even hot electrons might radiate inefficiently when the plasma is rarefied (Quataert \& Narayan 1998).

\subsection{The rate of spherical accretion according to Bondi}
The Bondi formula can be explained as follows. At a distance $r\sim R_A$ from the massive object, the gravitational energy of the gas particles is comparable to the thermal energy. It is plausible that all these particles fall onto the black hole. Further assume that at  $r\sim R_A$ the radial velocity is $\sim c_s$ and the density is $\sim \rho $. We then get the mass accretion rate given by (1). 

Is the pressure build-up on the way into the massive object important? Can the pressure forces counter the gravitational attraction? We can answer this question by first neglecting the pressure forces at small radii, $\ll R_A$. Then the radial velocity is equal to the free fall velocity, $\propto r^{-1/2}$. From mass continuity, the density is $\propto r^{-3/2}$. The pressure is $\propto r^{-3\gamma /2}$, where $\gamma$ is the adiabatic index of the gas. The acceleration from the pressure gradient is $\propto r^{-(3\gamma -1)/2}$. The gravitational acceleration is $\propto r^{-2}$. So long as $\gamma <5/3$, the pressure is not able to resist gravity. 

This qualitative derivation explains why the Bondi formula should work well if the accreting gas has an effective adiabatic index $\gamma _{\rm eff}\le 5/3$. In this case, the pressure forces cannot impede accretion, and the gas from $r\sim R_A$ must accrete at about the Bondi rate (1). Numerical simulations of a non-spherical accretion confirm the Bondi formula (Coker \& Melia 1997).

\subsection{$\gamma _{\rm eff}>5/3$}
Suppose that the accreting matter is a collisionless plasma. Suppose that radiation is negligible. We will show that heat conduction increases the effective adiabatic index to $\gamma _{\rm eff}>5/3$. Qualitatively, the heat conduction reduces the loss of thermal energy through the event horizon. The temperature of the flow increases faster than for an adiabatic plasma, which means that $\gamma _{\rm eff}>5/3$.

Quantitatively, $TdS=dQ$, or
\begin{equation}
-\rho Tv\partial _rs=r^{-2}\partial _r (r^2 \kappa \partial _r T).
\end{equation}
Here $\rho$ is the density, $T$ is the temperature, $v$ is the inflow (positive) velocity, $s$ is the entropy per unit mass, $\kappa$ is the thermal conductivity. We assume that the thermal conductivity $\kappa$ is given by a ``Shakura-Sunayev-like'' formula
\begin{equation} 
\kappa =m^{-1}\alpha \rho rv,
\end{equation}
where $\alpha \sim 1$ is a dimensionless positive constant, $m$ is the molecular mass. The Shakura-Sunayev-like prescription is plausible because the characteristic size and frequency of the turbulent structures should be $\sim r$ and $\sim v/r$. The thermal conductivity might be suppressed for a supersonic inflow, but our accretion flow is subsonic. 

With $s=m^{-1}{\rm ln}(T^{3/2}/\rho )$, equation (3) gives
\begin{equation} 
\rho Tv\left( {3\over 2}{T'\over T}-{\rho '\over \rho }\right) + \alpha r^{-2}(r^3\rho v T')'=0,
\end{equation}
where the prime denotes the r-derivative. For a stationary flow $\rho vr^2={\rm const}$, and
\begin{equation} 
T\left( {3\over 2}{T'\over T}-{\rho '\over \rho }\right) + \alpha (rT')'=0.
\end{equation}
If $\alpha$ is small, we can use the adiabatic relationship in the last term, $rT'=-T$. Then equation (6) gives 
\begin{equation} 
T^{{3\over 2}-\alpha}\propto \rho,
\end{equation}
that is 
\begin{equation} 
\gamma _{\rm eff}={5-2\alpha \over 3-2\alpha}.
\end{equation}

Since $\gamma _{\rm eff}>5/3$, pressure will impede the accretion. We now estimate the suppressed accretion rate. Assume that at the event horizon the temperature is close to $mc^2$. The density at the hole $\rho _h$ is given by the adiabatic law
\begin{equation} 
{\rho _h\over \rho }\sim \left( {c^2\over c_s^2}\right) ^{1\over \gamma _{\rm eff} -1}\sim \left( {R_A\over R_S}\right) ^{1\over \gamma _{\rm eff} -1}.
\end{equation}
The rate of accretion is $\dot{M} \sim R_S^2c\rho _h$, and from (8) and (9) we get (2).

\subsection{A solvable model}  
In this section we rigorously derive the accretion rate when heat conduction is important, equation (2). We assume a spherically symmetrical gas inflow. The radiative cooling is neglected. The flow is taken to be inviscid but non-adiabatic, i.e. we include thermal conduction and neglect viscosities. 

In terms of suppressing the rate of accretion, heat conduction is more important than non-sphericity or viscosity. Non-spherical accretion is known to proceed at about the Bondi rate (Coker \& Melia 1997). Viscosity suppresses the accretion rate, but only weakly, because when the inflow velocity decreases, the effects of viscosity decrease faster than the adiabatic effects (see also Turolla \& Nobili 1989).   

We treat the flow using a fluid approximation, but we should remember that in reality we are dealing with magnetized collisionless plasmas. However, the fluid approximation obeys the correct set of conservation laws, and should be close to reality. The qualitative conclusions of our analysis should be taken seriously. 

Consider a non-relativistic radial inflow of the $\gamma =5/3$ gas. The gas density $\rho$, the inflow (positive) velocity $v$, and temperature $T$ depend on the radial coordinate $r$ only, and satisfy the stationary equations of continuity, Euler's, and the thermal conduction.
\begin{equation} 
\dot{M}=4\pi r^2\rho v,
\end{equation}
\begin{equation} 
vv'=-{(\rho T)'\over m\rho}-{GM\over r^2},
\end{equation}
\begin{equation} 
{\rho Tv\over m}\left( {\rho '\over \rho }-{3\over 2}{T'\over T}\right) = r^{-2}(\kappa r^2T')'.
\end{equation}
We will use dimensionless units with $GM=1$, and the molecular mass $m =1$. 

The boundary condition at the surface of the compact object is unknown, but it is irrelevant. For concreteness, assume $T+\beta rT'=0$ at $r=R_S$, where $\beta$ is a dimensionless constant. At $r=\infty$, we assume $T=1$, so that $R_A\sim 1$. The boundary condition for $\rho$ is also irrelevant, because the system of equations is invariant under $\rho \rightarrow \lambda \rho$, $\dot{M} \rightarrow \lambda \dot{M}$. 

Our aim is to find the maximal possible value of the dimensionless mass accretion rate $\dot{m}\equiv \dot{M}/\rho (\infty )$, for which a smooth solution of the system exists. The ratio $\dot{M}/\dot{M}_{\rm Bondi}$ is equal to this maximal value. Obviously $\dot{m}\sim 1$ if $R_S\sim 1$. We need to find out how $\dot{m}$ depends on $R_S$ when $R_S\ll 1$. 

Scale out $\dot{M}$ by $\rho \rightarrow \dot{M}\rho/4\pi $, insert (4) and (10) into (12):
\begin{equation} 
\rho vr^2=1,
\end{equation}
\begin{equation} 
vv'=-\rho ^{-1}(\rho T)'-r^{-2},
\end{equation}
\begin{equation} 
T\left( {\rho '\over \rho }-{3\over 2}{T'\over T}\right) = \alpha (rT')'.
\end{equation}
Using (15), integrate (14) and obtain a system of two first-order differential equations
\begin{equation} 
{1\over 2} {1\over r^4\rho ^2}+{5\over 2}T+\alpha rT'={1\over r}+{5\over 2},
\end{equation}
\begin{equation} 
{\rho '\over \rho}\left( T-{1\over r^4\rho ^2}\right) =-{1\over r^2}+{2\over r^5\rho ^2}-T'.
\end{equation}
We need to find the dependence on $R_S\ll 1$ of the  maximal possible value of $\dot{m}\equiv 1/\rho (\infty )$ for which a smooth solution of the system exists. For small $r$, we can neglect $5/2$ in the right hand side of (16). Then, denoting $\tau=rT$, $f=(r^3\rho ^2)^{-1}$, $x=\log r$, we obtain 
\begin{equation} 
2\alpha {d\tau \over dx}=2-(5-2\alpha)\tau -f,
\end{equation}
\begin{equation} 
{df \over dx}=f{2+2d\tau /dx-5\tau -f\over \tau -f}.
\end{equation}
This system has a stable equilibrium point $\tau =2(5-2\alpha)^{-1}$, $f=0$ (it is assumed that $\alpha <5/2$). In the vicinity of this point, 
\begin{equation} 
{df \over dx}=-2\alpha f.
\end{equation} 
The system (18), (19) is an accurate approximation only for $r\ll 1$, i. e., for an x-duration $\delta x\sim \log (1/R_S)$. The quantity $f$ decreases according to (20) for an x-duration $\delta x=\log (1/R_S)-{\rm const}$. Therefore the final density is proportional to $(R_S)^{-\alpha }$. Since $\dot{m}\equiv 1/\rho (\infty )$, the scaling law (2) is proven. To check the answer, we integrated equations (16), (17) numerically.

\section{The Density and Temperature of the Collisionless Halo}

The standard Bondi solution and the new solution of \S 3 predict different densities near the black hole. In both cases the temperature is close to virial
\begin{equation}
T(r)\sim m_pc^2{R_S\over r}
\end{equation}
The density is given by the adiabatic. For the Bondi solution
\begin{equation}
n(r)\sim n_{\infty }\left( {R_A\over r}\right) ^{3/2},
\end{equation}
where $n_{\infty }$ is the density at a large distance from the hole (at accretion radius). When heat conduction is important, we use the effective value of the adiabatic index. This gives
\begin{equation}
n(r)\sim n_{\infty }\left( {R_A\over r}\right) ^{3/2-\alpha }.
\end{equation}

For Sgr A$^*$, under the assumption of \S 3, the new density is two orders of magnitude smaller. The density profile should, in principle, be observable. The deviations from the $r^{-3/2}$ law will prove that collsionless halos around black holes are not ideal.

\section{Discussion}
The low luminosity of the supermassive black holes existing in the centers of most galaxies (Richstone 1998) is traditionally explained by the extremely low radiative efficiency of the spherical accretion. We offer an alternative explanation: these black holes accrete at a rate much smaller than the Bondi rate. 

We have explained that electron heating and the radiative efficiency of the collisionless halos around black holes should be non-negligible. Then the observed low luminosity of many supermassive black holes should be due to the small accretion rate. 

We generalized the Bondi formula by including thermal conductivity. Thermal conductivity heats the flow, by transporting heat from the event horizon outward into the plasma. As the result of heat conduction, the pressure rises and impedes the accretion flow.

\acknowledgements This work was supported by NSF PHY-9513835. I thank Eliot Quataert, John Bahcall, and Martin Rees for useful discussions.

\end{document}